\documentclass[aps,prl,twocolumn,groupedaddress,showpacs,showkeys]{revtex4}

\usepackage{graphicx}

\begin{document}

\newcommand{\be}{\begin{equation}}
\newcommand{\ee}{\end{equation}}

\title{Observing the Casimir-Lifshitz Force Out of Thermal Equilibrium}

\date{\today}

\author{Giuseppe Bimonte}
\affiliation{Dipartimento di  Fisica, Universit{\`a} di Napoli Federico II, Complesso Universitario
MSA, Via Cintia, I-80126 Napoli, Italy}
\affiliation{INFN Sezione di
Napoli, I-80126 Napoli, Italy }

\begin{abstract}

The thermal Casimir-Lifshitz force among two bodies held at different temperatures displays striking features that are absent in systems in thermal equilibrium. The manifestation of this  force has been observed so far only  in Bose-Einstein condensates close to a heated substrate, but never between two macroscopic bodies.  Observation of the thermal Casimir-Lifhitz force out of thermal equilibrium with conventional Casimir setups is very difficult, because for  experimentally accessible separations the thermal force is small compared to the zero-temperature quantum Casimir force, unless prohibitively large temperature differences among the plates are considered.     We describe an apparatus that allows for a direct observation of the thermal force out of equilibrium for submicron separations and for moderate temperature differences between the plates.     
  
\end{abstract}



\maketitle

\author{ Giuseppe Bimonte}
 
Casimir-Lifshitz forces \cite{parse,book2},  i.e.  dispersion forces between polarizable bodies originating from quantum and thermal fluctuations of the electromagnetic (em) field, play an important role in different fields of science (physics, biology, chemistry) and in technology.  The first comprehensive theory of dispersion forces was developed in the fifties of last century  by Lifshitz \cite{lifs}, on the basis of Rytov's theory of electromagnetic fluctuations \cite{rytov}. Still today, Lifshitz theory is routinely used to interpret  experiments on dispersion forces.  

In its original formulation, Lifsihtz theory dealt  with two material slabs in thermal equilibrium. Recently, the theory has been generalized by the Trento group \cite{pita1,pita2,pita3} to situations out of equilibrium, in which the interacting bodies may have different temperatures.  The study of the thermal component of  the  Casimir-Lifshitz force has attracted much interest in recent years.
Observing the thermal force is very difficult, as  it becomes visible only at distances of the order of the thermal wave length $\lambda_T=\hbar c /k_B T$ (about 7 microns at room temperature).  At such large distances both the quantum Casimir force and the thermal force are very small, and thus very difficult to measure.  On the other hand, for smaller distances the thermal force is masked by the much stronger $T=0$ quantum component of the Casimir-Lifshitz force, and therefore it is difficult to separate it unambiguously.  As of now,   only two experiments have observed the thermal Casimir-Lifshitz force. The first one is the JILA experiment \cite{JILA}, which observed the thermal Casimir-Polder force between an ultracold atomic cloud placed at a distance of a few microns from a dielectric substrate. In order to enhance the thermal force, the measurement was done out of thermal equilibrium by heating the substrate, and was found to be in agreement with the theory developed in \cite{pita1}.  The second experiment by the Yale group \cite{lamorth} observed the equilibrium thermal Casimir force between a large Au sphere and a Au plate, in the wide range of separations form 0.7 to 7.3 $\mu$m. The theoretical interpretation of the Yale experiment is controversial \cite{critiz}, because of the presence in the signal of a ten times larger force of unclear origin, that was attributed to large electrostatic patches on the gold surfaces.          
 
Out of thermal equilibrium the Casimir-Lifshitz force displays remarkable features that disappear when the system is brought in a state of thermal equilibrium  \cite{pita1,pita2,pita3}. These features originate from a peculiar contribution ${\bar F}^{(\rm neq)}(T_1,T_2)$ to the non-equilibrium force, which is {\it antisymmetric} under an exchange of the bodies temperatures $T_1$ and $T_2$. The presence of 
such a term, first pointed out in \cite{pita1}  for the case of a polarizable small particle  in front of a flat dielectric surface, and then in \cite{pita2} for two plane-parallel slabs,  has been later shown to be a general feature of the non-equilibrium force between  two bodies of any shape and composition \cite{bimontescat,antezza1,antezza2,krugertrace}. Being antisymmetric in the bodies temperatures, this term can have either sign and it can be harnessed to tune the force both in strength and sign  \cite{dilut}, and to realize self-propelling systems \cite{krugerself}.  This thermal force enjoys more striking features: it vanishes identically for two bodies with identical scattering matrices \cite{pita3,bimontescat,antezza1,antezza2,krugertrace}, and it is non-additive in the limit where one of the two bodies is a rarefied gas \cite{pita1,pita2}.
In view of its unique features, it would be clearly of great interest to observe the effect of this term in the Casimir force between two {\it macroscopic} bodies held at different temperatures. So far this has been an impossible task, because  in order to observe this term by  current  Casimir setups it would be necessary to achieve a large temperature difference between the plates (hundreds of degrees), and to go to separations of several microns.  Both things are very difficult to realize in practice. 

\begin{figure}[h]
\includegraphics[width=.9\columnwidth]{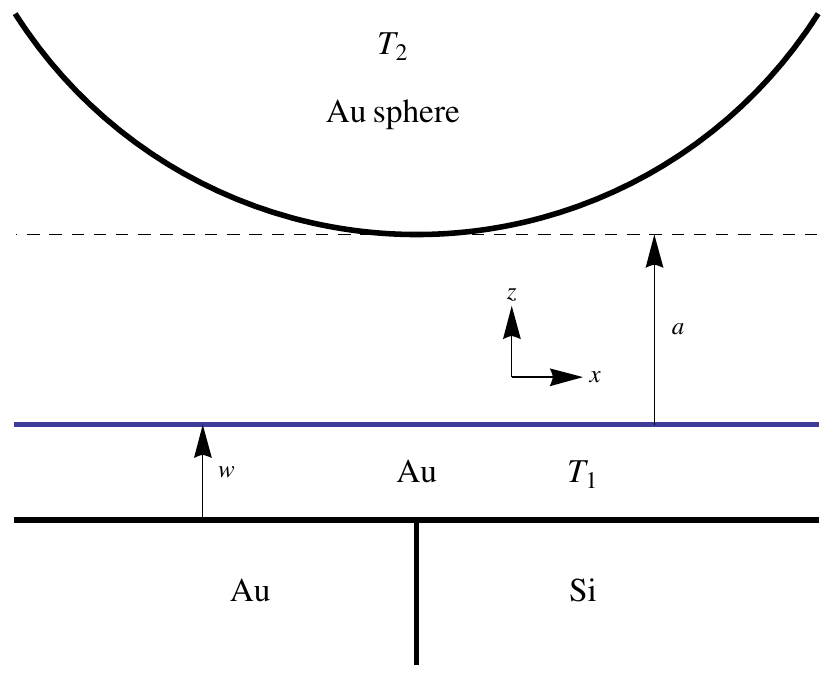}
\caption{ The setup consists of a  gold sphere at temperature $T_2$   above a  planar slab at temperature $T_1$.  The planar slab is divided in two regions  respectively made of gold and (high resistivity) silicon, and is fully  covered  with a plane-parallel gold over-layer of uniform thickness $w=100$ nm.   }
\label{fig1}
\end{figure}

In this Letter we describe an apparatus that should allow for a direct observation of the antisymmetric component of the non-equilibrium thermal Casimir-Lifshitz force at submicron distances, with small temperature differences between the plates.  The scheme is  based on a {\it differential} force measurement, similarly to  two setups recently proposed by the author \cite{hide1,hide2} to probe the {\it equilibrium} thermal Casimir force between magnetic and non-magnetic plates.     The setup, schematically shown in Fig.\ref{fig1},   consists of a gold sphere  of radius $R$ at temperature $T_2$  placed at a (minimum) distance $a$ from  a planar slab at temperature $T_1$, divided in two regions  made of   gold and of (high resistivity) silicon respectively. 
The key feature of the  apparatus is the gold over-layer of thickness $w$, covering both the gold and the silicon regions of the plate. 
For any fixed sphere-plate separation $a$, we consider measuring  the {\it difference} 
\be\Delta F (T_1,T_2)=F_{\rm Si}(T_1,T_2)-F_{\rm Au}(T_1,T_2)\ee among  the values $F_{\rm Au}(T_1,T_2)$ and $F_{\rm Si}(T_1,T_2)$  of  the (normal) Casimir force on the sphere   (negative forces correspond to attraction towards the plate) that obtain when the tip of the sphere is respectively above a point $q$   deep in the Au region, and a point $p$   deep in the Si region \footnote{Out of thermal equilibrium, the  force  difference $\Delta F (T_1,T_2)$ includes in general an uninteresting distance-independent contribution originating from the thermal radiation of the environment, that we consider substracted from the signal. For a detailed discussion of this contribution see \cite{pita3}. }. The principle behind this differential measurement can be easily explained. One considers that  em quantum fluctuations contributing to the $T=0$ Casimir force have characteristic frequencies of the order of $\omega_c=c/(2 a)=5 \times 10^{14}$ rad/s, for a separation of 300 nm. Photons with this frequency have a penetration depth $\delta_0$ in Au of 20 nm, or so. On the other hand, inspection of the spectrum of the thermal Casimir-Lifshitz force (and in particular of the antisymmetric contribution ${\bar F}^{(\rm neq)}(T_1,T_2)$   that is our main interest)  reveals that the important photon frequencies are smaller than $0.05 \times k_B T/\hbar \simeq 2 \times 10^{12}$ rad/s  for temperatures $T$ around 300 K. The penetration depth $\delta_T$ of these thermal photons in Au is  around 160 nm.  Therefore, if the thickness $w$ of the gold over-layer is chosen such that
\be
\delta_0 \ll w \ll\delta_T\;,\label{rangew}
\ee 
it is clear that the over-layer filters out from the signal $\Delta F$ the uninteresting $T=0$ component of the Casimir force, which would otherwise mask the much weaker thermal force. On the contrary, low frequency thermal photons contributing to ${\bar F}^{(\rm neq)}(T_1,T_2)$,  being able to traverse the Au over-layer, are sensitive to the different optical properties of the Au-Si substrates.  In our computations  we took $w=100$ nm, and we found that for a large sphere radius $R \gg a$  the signal $\Delta F(a)$ is essentially equal (as it will be better explained in the sequel of the paper) to the antisymmetric component ${\bar F}^{(\rm neq)}_{\rm Si}(T_1,T_2)$ of the thermal force  that obtains when the sphere is above the Si sector:
\be
\Delta F(T_1,T_2) \simeq {\bar F}^{(\rm neq)}_{\rm Si}(T_1,T_2)\;. \label{filter}
\ee    
This shows that by our measurement scheme  it is possible to directly observe the non-equilibrium thermal force discovered by the Trento group.  Another important virtue of the proposed scheme is  that it is  immune by design from the problem of electrostatic patch forces that represent a major difficulty in conventional Casimir absolute force   measurements \cite{onof,onof2,onof3,decca,iannuzzi2,dalvit2}. This is so because the patch electrostatic force is independent of the position of the sphere tip above the Au over-layer, and therefore it cancels out from $\Delta F$. In this regards our scheme  is similar to recent  differential Casimir-less  experiments  by the IUPUI group \cite{deccaiso,decca7} searching for  non-newtonian gravity in the submicron range, which also utilized Au over-layers of thicknesses similar to ours to screen out both electrostatic and Casimir forces. It has been estimated recently \cite{speake} that for separations larger than 200 nm  random fluctuations of the patch potential from point to point on the surface of the Au overlayer imply a limit  0.1 fN on the sensitivity of the IUPUI apparatus.

\begin{figure}[h]
\includegraphics[width=.9\columnwidth]{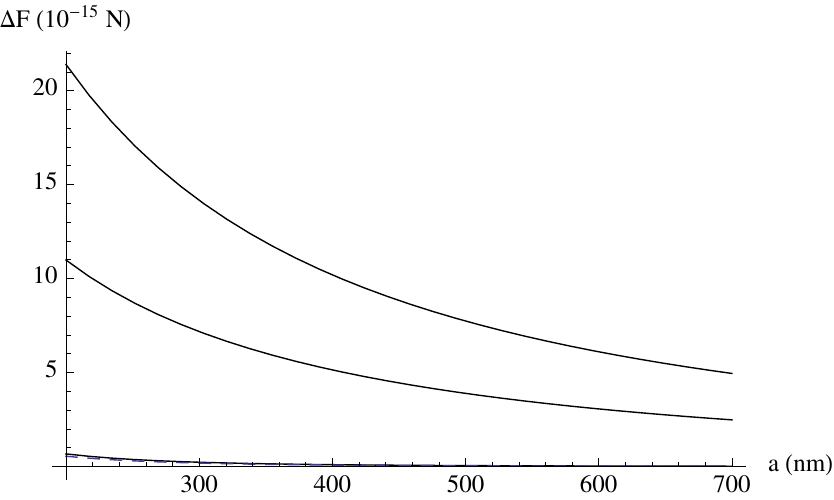}
\caption{Force difference $\Delta F$ (in fN) versus separation $a$ in nm, for a Au sphere of radius $R=150$ micron. The three solid curves, from top to bottom, correspond to  a fixed sphere temperature $T_2=300$ K, and to  three temperatures of the Au-Si plate $T_1=350$ K, 325 K and 300 K respectively. The dashed line represents the equilibrium force difference for $T_1=T_2=350$ K.}
\label{fig2}
\end{figure}

We turn now to the computation of $\Delta F$ for our apparatus. In \cite{pita3} it is shown that the Casimir pressure $F^{(\rm PP)}(T_1,T_2)$ between two plane-parallel plates at different temperatures is equal to the average of the equilibrium Casimir pressures corresponding to the two temperatures, plus an extra term ${\bar F}^{\rm (neq)}(T_1,T_2)$ which is antisymmetric in the temperatures. We shall see below that $\Delta F$ has a similar structure.  To make the computation simple, we make two assumptions. First, we assume that both points $p$ and $q$ at which the  force is measured are at a horizontal distance $s$ from the Au-Si boundary of the plate, which is much larger than the typical radius $\rho=\sqrt{R a}$ of the circular region around the sphere tip which contributes significantly to the Casimir force.   The force $F_{\rm Au}(T_1,T_2)$  can then be  identified with the force ${\tilde F}_{\rm Au}(T_1,T_2)$  between a gold sphere  and a homogenous gold plate, while $F_{\rm Si}(T_1,T_2)$ becomes identical to the force  ${\tilde F}_{\rm Si}(T_1,T_2)$ between the same gold sphere and a two-layer plane-parallel slab, consisting of a gold layer of thickness $w$ deposited over a uniform Si slab. We further assume, as it is usually the case in Casimir experiments, that the sphere radius $R$ be much larger than the separation $a$, $R \gg a$. For $R \gg a$, we can use the   proximity force approximation (PFA) \cite{parse,book2} to estimate both ${\tilde F}_{\rm Au}(T_1,T_2)$ and ${\tilde F}_{\rm Si}(T_1,T_2)$. According to the PFA  the force $F^{\rm (sp)}$ between a large sphere and a plate can be expressed in terms of  the potential ${\cal U}^{(\rm PP)}$  for the unit-area force $F^{\rm (PP)}=-\partial \,{\cal U}^{\rm (PP)}/\partial \,a$ of the corresponding plane-parallel system:
\be
{F}^{\rm (sp)}= 2 \pi \,R \,{\cal U}^{(\rm PP)}\;.\label{PFA}
\ee  
The  PFA formula Eq. (\ref{PFA})  holds for any  short-range interaction between  gently curved surfaces,  and it is valid also for the Casimir force out of thermal equilibrium. 
The PFA has been widely used to interpret Casimir experiments \cite{book2}  (see  \cite{kruger} for more applications of the Proximity Approximation). 
It is now known  that the PFA represents  the leading term in a gradient expansion of the Casimir force, in powers of the slopes of the bounding surfaces \cite{fosco2,bimonte3,bimonte4}. 
By the PFA  Eq. (\ref{PFA})  one gets:
\be
\Delta F(T_1,T_2)= 2 \pi R \left[ {\cal U}_{\rm Si}^{(PP)}(T_1,T_2)-{\cal U}_{\rm Au}^{\rm (PP)}(T_1,T_2)\right]\;,\label{setupforce}
\ee 
where $ {\cal U}_{\rm Au}^{\rm (PP)}(T_1,T_2)$ is the Casimir potential for two Au slabs at temperatures $T_1$ and $T_2$ respectively, and $ {\cal U}_{\rm Si}^{\rm (PP)}(T_1,T_2)$ is the potential for a Au slab at temperature $T_2$ in front of a two-layer Au-Si slab at temperature $T_1$.  
The   potential ${\cal U}^{\rm (PP)}(T_1,T_2)$ for two plane parallel dielectric slabs at different temperatures can be found  easily by integrating the formula for the non-equilibrium unit-area force $F^{\rm (PP)}(T_1,T_2)$ provided  in \cite{pita3}:
\be
{\cal U}^{\rm (PP)}(T_1,T_2)=\frac{1}{2}\left[{\cal F}(T_1)+{\cal F}(T_2) \right]+{\bar {\cal U}}^{(\rm neq)}(T_1,T_2)\;.\label{noneqfor}
\ee
In this formula,   ${\cal F}(T)$ denotes the well-known Lifshitz formula for the equilibrium unit-area Casimir free energy:
$$
{\cal F}(T)=\frac{k_B T}{2 \pi}\sum_{l=0}^{\infty}\left(1-\frac{1}{2}\delta_{l0}\right)\int_0^{\infty} d k_{\perp} k_{\perp}  
$$
\be
\times \; \sum_{j={\rm TE,TM}} \log \left[1- {e^{-2 a q_l}}{R^{(1)}_{j}({\rm i} \xi_l,k_{\perp})\;R^{(2)}_{j}({\rm i} \xi_l,k_{\perp})} \right]\;,\label{lifs}
\ee
while ${\bar{\cal U}}^{(\rm neq)} (T_1,T_2) $ has the expression:
\begin{widetext}
$$
{\bar {\cal U}}^{(\rm neq)}(T_1,T_2)=\frac{\hbar}{4 \pi^2} \int_0^{\infty} d \omega [n(\omega,T_1)-n(\omega,T_2)]\int_{0}^{\infty} d k_{\perp} k_{\perp} \sum_{j={\rm TE,TM}} 
{\rm Im} \,[\log\,(1-e^{2 {\rm i} a k_z} R_j^{\rm (1)}R_j^{\rm (2)})] 
$$
\be
\times\;\left[\theta(\omega/c -k_{\perp})\frac{|R_j^{\rm (2)}|^2-|R_j^{\rm (1)}|^2}{1-|R_j^{\rm (1)}R_j^{\rm (2)}|^2}+\theta(k_{\perp}-\omega/c) \frac{{\rm Im}(R_j^{\rm (1)}R_j^{\rm (2)*})}{{\rm Im}(R_j^{\rm (1)}R_j^{\rm (2)})}\right]\;.\label{noneq}
 \ee
\end{widetext}
In the above Equations  $R_j^{\rm (1)},  R_j^{\rm (2)}$ denote the reflection coefficients of the slabs for polarization $j$,   $k_B$ is Boltzmann constant, $\xi_l=2 \pi l k_B T/\hbar$ are the (imaginary) Matsubara frequencies, $k_{\perp}$ is the modulus of the in-plane wave-vector, $q_l=\sqrt{\xi_l^2/c^2+k_{\perp}^2}$, $k_z=\sqrt{\omega^2/c^2-k_{\perp}^2}$, $\theta(x)$ is the unit step-function ($\theta(x)=0$ for $x<0$ and $\theta(x)=1$ for $x>0$),  and $n(\omega,T)=(\exp[\hbar \omega/(k_B T)]-1)^{-1}$ is the Bose-Einstein distribution. According to Eq. (\ref{noneqfor}),  out of equilibrium the potential ${ {\cal U}}^{\rm (PP)}(T_1,T_2)$ is equal to the average of the  equilibrium Casimir free energies ${\cal F}(T)$ at temperatures $T_1$ and $T_2$, plus a genuinely non equilibrium contribution  ${\bar {\cal U}}^{(\rm neq)}(T_1,T_2)$. The latter term is {\it antysimmetric} in the temperatures $T_1,T_2$, and {\it vanishes identically} if   the slabs have {\it identical} reflection coefficients. Remarkably, this structure of the Casimir-Lifshitz force out of thermal equilibrium has been shown to be valid  also for non-parallel plates of arbitrary shapes and constitution \cite{bimontescat,antezza1,antezza2,krugertrace}. 
In order to evaluate Eq. (\ref{setupforce}), one   substitutes in  Eqs. (\ref{lifs}-\ref{noneq}) the reflection coefficient $ R_j^{\rm (2)}$ by the reflection coefficient  $R^{(\rm Au)}_{j}$ of a Au slab, and $ R_j^{\rm (1)}$ by either $R^{(\rm Au)}_{j}$ or by  the reflection coefficient $R^{(\rm Si)}_{j}$
of a Si slab covered by a gold layer of thickness $w$.  The   reflection coefficient $R^{(\rm Au)}_{j}$ is equal to the Fresnel coefficient $r_j^{(0 \rm Au)}$  given in Eqs. (\ref{freTE} - \ref{freTM}) below with $a=0$, $b=$Au, while $R^{(\rm Si)}_{j}$  is provided by the following formula:
\be
R_{j}^{({\rm Si})}( \omega,k_{\perp})=\frac{r_{j}^{(0{\rm Au})}+e^{2 {\rm i}\,w\, k_z^{({\rm Au})}}\,r_{j}^{({\rm Au Si})}}{1+e^{2{\rm i}\,w\, k_z^{({\rm Au})}}\,r_{j}^{(0{\rm Au})}\,r_{j}^{({\rm Au Si})}}\;.
\ee   
Here $r^{(ab)}_{\alpha}$ are the Fresnel reflection coefficients for a planar interface separating medium a from medium b: 
\be
r^{(ab)}_{\rm TE}=\frac{ k_z^{(a)}- k_z^{(b)}}{ k_z^{(a)}+ k_z^{(b)}}\;,\label{freTE}
\ee
\be
r^{(ab)}_{\rm TM}=\frac{\epsilon_{b}(\omega) \,k_z^{(a)}-\epsilon_{a}(\omega) \,k_z^{(b)}}{\epsilon_{b}(\omega) \,k_z^{(a)}+\epsilon_{a}(\omega) \,k_z^{(b)}}\;,\label{freTM}
\ee
where
$ k_z^{(a)}=\sqrt{\epsilon_a(\omega)   \omega^2/c^2-k_{\perp}^2}\;$, 
 $\epsilon_a$   denotes the electric   permittivity of medium $a$, and we define $\epsilon_0=1$. In our computations, we used the tabulated optical data of Au and Si  \cite{palik}. The data of Au were extrapolated towards zero frequency via the  Drude model $\epsilon_{\rm Dr}=1-\omega_p^2/[\omega(\omega+{\rm i} \gamma)]$, with
$\omega_{p}=8.9 \,{\rm eV}/\hbar$,  $\gamma=0.035\, {\rm eV}/\hbar$ \footnote{For the temperatures that we consider, the temperature variation of $\gamma$ has a negligible effect.}. We are now in a position to better justify  Eq. (\ref{filter}), showing that  $\Delta F$ measures the non equilibrium thermal Casimir-Lifshitz force.  We remarked earlier that if thickness $w$ of the gold over-layer is chosen in the range in Eq.(\ref{rangew}), the $T=0$  component of the Casimir force, which is included in the first two terms on the r.h.s. of Eq. (\ref{noneqfor}), is filtered out from $\Delta F$. As to thermal component of the Casimir-Lifshitz force, a distinction has to be made among the thermal correction to the equilibrium force, which is again included in the first two terms of Eq. (\ref{noneqfor}), and the truly non-equilibrium contribution provided by the last term on the r.h.s. of Eq. (\ref{noneqfor}). The equilibrium thermal correction has a characteristic frequency of the order of the first Matsubara mode $\xi_1=2 \pi k_BT /\hbar=2.5 \times 10^{14}$ rad/s at room temperature. Since this radiation has a penetration  depth in Au $\delta_T^{(\rm eq)}\simeq 20\, {\rm nm} \ll w$,  it is clear that the equilibrium thermal component of the force is filtered out as well by the overlayer. The situation with the 
non-equilibrium force proportional to ${\bar {\cal U}}^{(\rm neq)}(T_1,T_2)$ is remarkably different.  When the sphere tip is above the gold sector of the plate this contribution is zero, because  ${\bar {\cal U}}^{(\rm neq)}(T_1,T_2)$ vanishes identically for two  surfaces made of the same material. When  the sphere tip is instead above the Si sector of the plate, the non-equilibrium contribution ${\bar {\cal U}}^{(\rm neq)}(T_1,T_2)$ is different from zero, and by inspection of its spectrum we estimated that for submicron separations it receives its main contribution from evenescent waves with TE polarization in the frequency range around   $0.05 \times k_B T/\hbar \simeq 2 \times 10^{12}$ rad/s  for temperatures $T$ around 300 K.  Since the penetration depth $\delta_T$ of such a radiation in Au,  around 160 nm, is much larger than $w=100$ nm, this contribution to the thermal force is strongly affected by the Au-Si interface.  The conclusion of all these considerations, fully confirmed by numerical computations, is that for our setup:
\be
\Delta F \simeq 2 \pi R   \,{\bar {\cal U}}^{(\rm neq)}_{\rm Si}(T_1,T_2)\;.\label{signal}
\ee
In Fig. \ref{fig2} we show a plot of $\Delta F$ (in fN) versus separation $a$ in nm, for a sphere radius $R=150$ micron and a Au overlayer of thickness $w=100$ nm.  The three solid curves, from top to bottom, correspond to  a fixed sphere temperature $T_2=300$ K, and to  three temperatures of the Au-Si plate $T_1=350$ K, 325 K and 300 K respectively. The plot displays also (dashed line) the equilibrium force difference for $T_1=T_2=350$ K. 
The close proximity of the two equilbrium curves confirms that the force difference $\Delta F$ seen for $T_1 \neq T_2$ arises entirely from the non-equilibrium thermal force proportional to ${\bar {\cal U}}^{(\rm neq)}_{\rm Si}(T_1,T_2)$, in accordance with Eq. (\ref{signal}).  
The isolectronic IUPUI Casimir-less experiments \cite{deccaiso,decca7} searching for non-newtonian gravity in the submicron range,  measured dynamically  the differential force bewteen a Au sphere glued to a microtorsional oscillator, and a rotating disk consisting of alternating Au and Si regions covered by a Au overlayer. A sensitivity better than 0.3 fN in force differences was reported in the separation range from 200 to 1000 nm, for an integration time of 3000 s. If this level of sensitivity can be preserved in the presence of a temperature difference between the sphere and the disk of a few tens of degrees, it should be easily possible to measure precisely the out-of-equilibrium thermal force displayed in Fig. \ref{fig2}. 
\begin{figure}[h]
\includegraphics[width=.9\columnwidth]{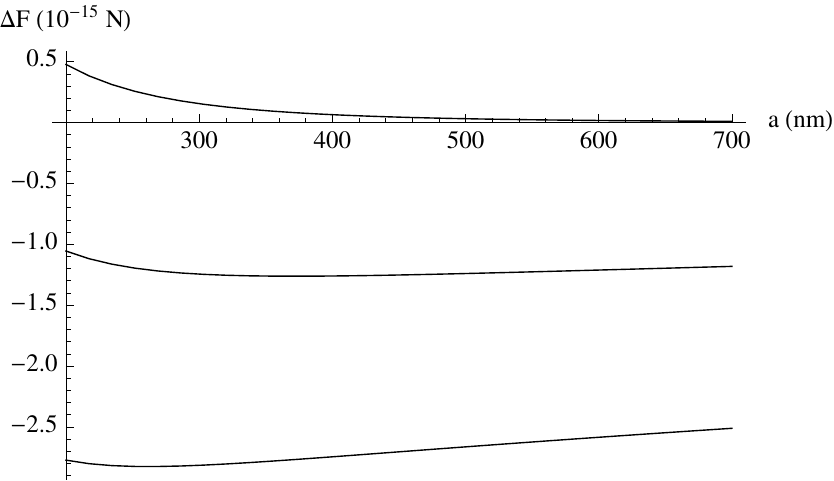}
\caption{Force difference $\Delta F$ (in fN) versus separation $a$ in nm, for a Au sphere of radius $R=150$ micron at a fixed temperature $T_2=300$ K. The three curves, from top to bottom, correspond to temperatures of the Au-Si plate equal to $T_1=300$ K, 325 K and 350 K, respectively. For this plot the conduction electrons of Au have been modelled  as a dissipantionless plasma.}
\label{fig3}
\end{figure}
 
An ongoing controversy in  Casimir physics concerns the influence of relaxation processes of conduction electrons on the thermal Casimir force.  
Surprisingly, several Casimir experiments  appear to be in agreement with Lifshitz theory only if conduction electrons are modelled by the dissipantionless plasma model of infra-red optics, while inclusion of dissipation via the plausible Drude model results in predictions of the Casimir force that are inconsistent with the data.   
We address the reader to the book \cite{book2} for a discussion of this subtle problem.  In Fig. \ref{fig3}  we show the plasma model prediction of $\Delta F$ (in fN) versus separation $a$ (in nm)
computed for the same temperatures $T_1$ and $T_2$ as in Fig. \ref{fig2}.  Contrary to  the Drude model result shown in Fig. \ref{fig2}, the plasma model predicts that $\Delta F$ should shift towards negative values as the temperature of the plate is increased.  The opposite behaviors of $\Delta F$ predicted by the two prescriptions should be easily detectable.      It thus appears that observation of the non-equilibrium thermal force by the present apparatus  should allow for a definitive experimental   resolution of the problem of dissipation in the Casimir effect.  

We have described an apparatus by which it should be possible to observe for the first time the thermal Casimir-Lifshitz force between two macroscopic surfaces out of thermal equilibrium. The sensitivity achieved by recent isolectronic Casimir-less expriments should allow for a precise measurement of the thermal Casimir force in the submicron region, and for moderate temperature differences between the plates. Apart from shedding light on the elusive thermal Casimir force, such an experiment might also allow to resolve a long-standing controversy regarding the role of dissipation in the Casimir effect.

\end{document}